\documentclass[conference]{IEEEtran}
\IEEEoverridecommandlockouts
\usepackage{cite}
\usepackage{amsmath,amssymb,amsfonts}
\usepackage{algorithmic}
\usepackage{graphicx}
\usepackage{textcomp}
\usepackage{xcolor}
\usepackage{setspace}
\def\BibTeX{{\rm B\kern-.05em{\sc i\kern-.025em b}\kern-.08em
		T\kern-.1667em\lower.7ex\hbox{E}\kern-.125emX}}
\usepackage{subfigure}
\newtheorem{my_theorem}{Theorem}

\newcommand*{\J}{\jmath}%
\usepackage{enumitem}

\title{ Performance Analysis of Cooperative Relaying for Multi-Antenna RF Transmissions over THz Wireless Link}

\author{
	\IEEEauthorblockN{Pranay Bhardwaj and S. M. Zafaruddin}\\
	\IEEEauthorblockA{ Deptt. of Electrical and Electronics Engineering, 
		BITS Pilani, Pilani-333031, Rajasthan, India.\\ Email: \{p20200026, syed.zafaruddin\}@pilani.bits-pilani.ac.in}
	\thanks{This work was supported in part by the Science and Engineering Research Board (SERB), India under MATRICS Grant MTR/2021/000890 and Start-up Research Grant SRG/2019/002345.}
}

\begin{document}
	\maketitle 
\begin{abstract}
Recent research has focused on single antenna radio-frequency (RF) and terahertz (THz) wireless systems to mix the access link with the backhaul. In this paper, we evaluate the performance of a mixed RF-THz system employing multiple antenna-assisted access point (AP) for the RF link and single-antenna THz transmissions. We employ an equal gain combining (EGC) receiver at the AP and use the fixed-gain amplify and forward (AF) relaying protocol to interface the RF and THz links. We derive analytical expressions for probability density function (PDF) and cumulative distribution function (CDF) of the end-to-end SNR for the considered system assuming  independent and non-identically distributed (i.ni.d.)  $\alpha$-$\mu$ distribution to model for both RF and THz channels and pointing errors in the THz link. We analyze the system performance using the outage probability, average bit error rate (BER), and ergodic capacity involving bivariate Fox's H-function. We use the  residue method to develop asymptotic analysis using Gamma functions to show the impact of the various channel and system parameters on the outage probability and average BER in the high SNR regime. We use computer simulations to depict the scaling of the performance with an increase in the number of antennas at the AP for signal reception in the access link.
\end{abstract}
		

\section{Introduction}
Terahertz (THz) wireless is emerging as a potential alternative for backhaul links to support fiber-like high-speed connectivity for beyond 5G and 6G networks  \cite{Elayan_2019, Koenig_2013_nature}. Deployment of optical fiber links may not be feasible everywhere, for example, in difficult terrains to provide ubiquitous wireless connectivity. Signal transmission in the THz band is less susceptible to atmospheric weather conditions such as fog and atmospheric turbulence due to the scintillating effect of signal propagation, as observed in free-space optical (FSO) communications.  However, the THz band suffers from higher path loss due to molecular absorption \cite{Kokkoniemi_2018} in addition to the signal fading and misalignment errors between the transmitter and receiver.    Radio frequency (RF)  is a matured technology for the access link in contemporary wireless communications. It is desirable to analyze the mixed wireless link consisting of THz backhaul and RF technology for access network in the next-generation wireless systems providing seamless, ultra high speed, and reliable connectivity.
 
Recently, there has been  extensive research on the  mixed  RF-THz  \cite{Boulogeorgos_Error,Pranay_2021_TVT,Pranay_2021_Letters} and RF-FSO systems (see \cite{Lee_2011_FSO_RF,Ashrafzadeh_2019_FSO_RF} and references therein). These hybrid network architectures can achieve the desired quality of communication for the next generation wireless systems and have the advantage of backward compatibility with the existing infrastructure. The authors in \cite{Boulogeorgos_Error} considered a mixed THz-RF system using the decode-and-forward (DF) relaying protocol to facilitate communication in the heterogeneous network and developed outage probability and average bit-error-rate (BER) performance assuming identical and independent (i.i.d.) $\alpha$-$\mu$ channel fading for both RF and THz links. In \cite{Pranay_2021_TVT}, we analyzed the THz-RF transmission using the outage probability, average BER, and ergodic capacity for several scenarios of practical interest, considering a more realistic asymmetrical fading for both the links. We also analyzed the mixed RF-THz system using the fixed-gain amplify-and-forward (AF) relay under the effect of generalized pointing errors and $\alpha$-$\mu$ fading for THz and $\alpha$-$\kappa$-$\mu$ shadowed fading channel for the RF \cite{Pranay_2021_Letters}. It should be noted that the $\alpha$-$\mu$ fading model is experimentally validated distribution to model the short-term fading for THz band as validated through recent experimental results \cite{Papasotiriou2021}. In the aforementioned and related literature, a single-antenna equipped access point (AP) or base station (BS) has been considered for the access link over RF.

The performance of the RF-THz link may be limited by the weaker of two links, mostly  RF transmissions under deep fading scenarios caused by non-line-of-sight (NLOS) multi-path propagation and shadowing effect in a cluttered access network. Hence, it is desirable to improve the performance of the RF link to match with the THz  to experience the actual RF-THz performance. The use of multiple antennas at the AP/BS is a standard configuration to harness the spatial diversity for RF signal reception and seems to be a viable alternative to enhance the access link performance. Note that using multiple antennas at the AP may provide another design criteria for the RF-THz based hybrid network. The mixed RF-THz with multi-antenna RF signaling has not yet been investigated to the best of the author's knowledge.

This paper evaluates the performance of a dual-hop RF-THz uplink transmission with multiple antenna-equipped AP/BS for RF signal reception. We employ equal gain combining (EGC) at the multi-antenna receiver and use the fixed-gain AF relaying for the mixed RF-THz system. We develop statistical results for the considered system by deriving analytical expressions of probability density function (PDF) and cumulative distribution function (CDF) for the end-to-end SNR assuming independent and non-identically distributed (i.ni.d.)  $\alpha$-$\mu$ distribution for both RF and THz channels and zero-boresight pointing errors in the THz link. We analyze the outage probability, average bit error rate (BER), and ergodic capacity performance involving bivariate Fox's H-function. We use the  residue method to represent the asymptotic analysis of the outage probability and average BER  using Gamma functions depicting the impact of various system parameters in the high SNR regime. We validate our analytical expressions using Monte Carlo analysis and use computer simulations to demonstrate the scaling of the RF-THz performance with an increase in the number of antennas at the AP.

\emph{Notations}: In the following, the superscripts $ {}^{\rm (R)} $ and  $ {}^{\rm T} $ denote the signal at the relay
and transpose operation, respectively. $(\cdot)_r$ denotes  parameters of the RF link from the source to relay, and $(\cdot)_t$ denotes the parameter of the THz link from the relay to the destination. $\Gamma(.)$, $G_{p,q}^{m,n}\big(.|.\big)$, and $ H_{p,q}^{m,n}\big(.|.\big) $  represent the Gamma function, the Meijer's G-function, and the Fox's H-function \cite{Mathai_2010}, respectively. The imaginary number is denoted by $\J$.
\begin{figure}	
   	\includegraphics[width=\columnwidth]{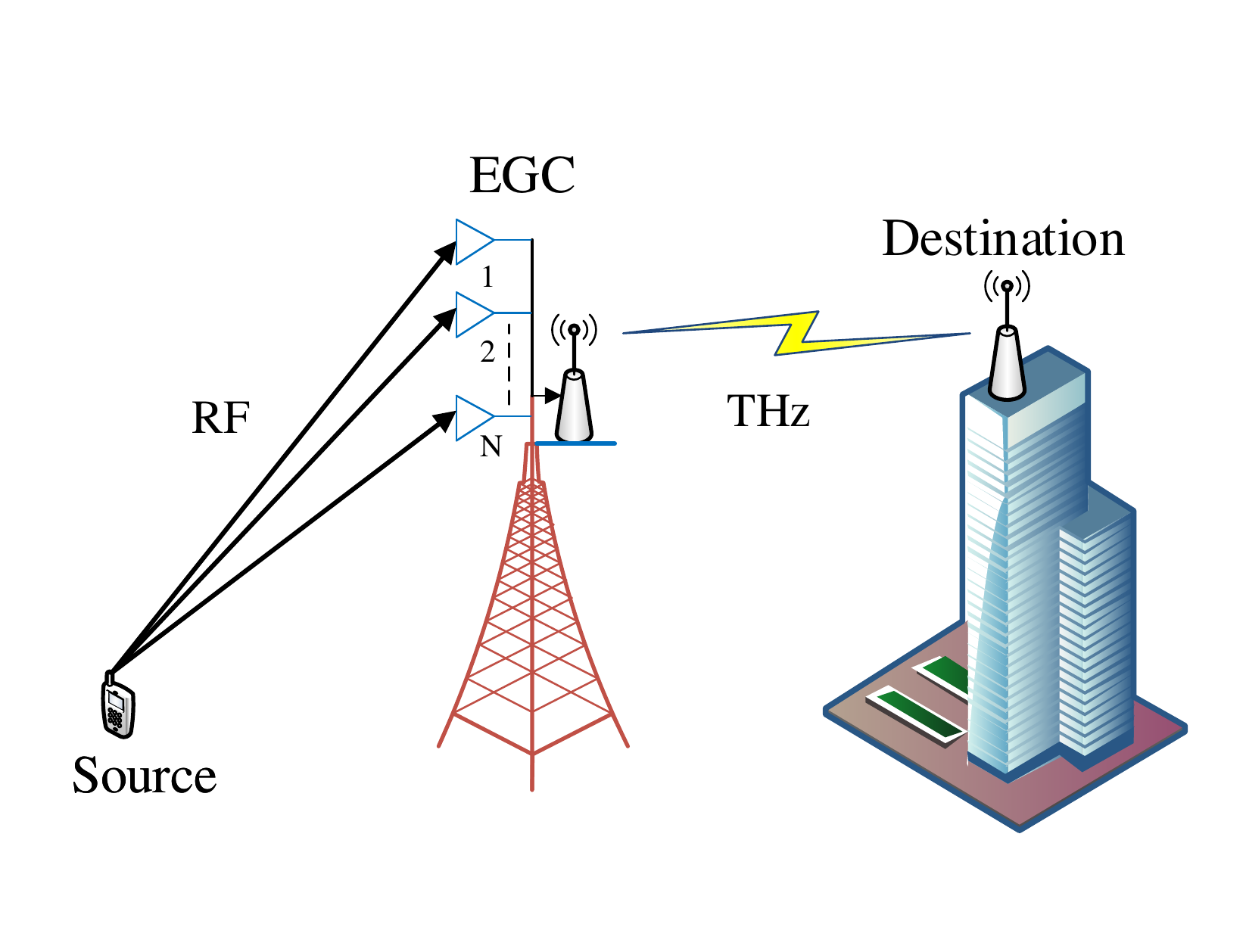}
 	\vspace{-6mm}
 	\caption{Schematic digram for the mixed link with EGC-enabled multi-antenna RF and THz.}
 		\label{system_model}
\end{figure} 
\section{System Model}\label{sec:system_model}
A single-antenna user in the access network wishes to communicate with a destination through multiple antennas ($N\geq 1$) equipped AP/BS, as shown in Fig.~\ref{system_model}. The signal transmission from the user to the relay occurs in the RF band (typically in the 6GHz band) and relay to the destination in the THz band (at a carrier frequency of  $275$ \mbox{GHz}).A frequency converter is employed to facilitate changing carrier frequency at the relay. 

In the first hop, the received signal vector $\mathbf{y}^{(R)}$ at the relay can expressed as
\begin{eqnarray} 
  \mathbf{y}^{(R)} = H_{r}  \mathbf{h_{r}}s + w_r,
\end{eqnarray}
where  is $\mathbf{y}^{(R)}= [Y_{1}^{(R)}, Y_{2}^{(R)},\cdots, Y_{N}^{R}]^{\rm T} $ with $Y_{i}^{(R)}$ denoting the received signal at the $i$-th antenna,   $s$ is the transmitted signal from the source  with power $P$, $w_r$ is the additive white Gaussian noise (AWGN) signal with variance $\sigma_{w_r}^2$, $H_{r}$ is the deterministic  channel path-gain, and $\mathbf{h_{r}} = [h_{1}, h_{2},\cdots, h_{N}]^{\rm T} $with $h_{i}$ denoting the   fading coefficient from the source to the  $i$-th antenna at the relay. 

We use the EGC receiver at the relay such that received signals of $N$ antennas are co-phased and added to get resultant SNR for the RF link $\gamma_r$ \cite{Annamalai_2000_EGC}:
\begin{eqnarray} \label{eq:egc_snr_eqn}
\gamma_r = \frac{1}{N} \Big(\sum_{i=1}^{N} \sqrt{\gamma^{(R)}_i}\Big)^2
\end{eqnarray} 
where $ \gamma^{(R)}_i=\bar{\gamma}_{\rm rf}|h_i|^2$ denotes the  SNR at the $i$-th antenna and $\bar{\gamma}_{\rm rf}= \frac{P |H_{r}|^2}{\sigma_{w_r}^2}$ is the average SNR for the RF link.  We model the short term fading  $|h_{i}|$ using the generalized  $\alpha$-$\mu$ distribution and use the transformation of random variables  to get PDF of the SNR for the $i$-th antenna as
\begin{equation} \label{eq:pdf_hf_rf}
	f_{|\gamma_{i}^{R}|}(x) = \frac{A_i \gamma^{\frac{\alpha_i\mu_i}{2}-1}}{2 \bar{\gamma}_{\rm rf}^{\frac{\alpha_i\mu_i}{2}}} \exp \bigg(- B_i {\Big(\sqrt{\frac{\gamma}{\bar{\gamma}}_{i}}\Big)^{\alpha_i}}\bigg)
\end{equation}
where  $A_i = \frac{\alpha_i\mu_i^{\mu_i}}{\Omega_i^{\alpha_i\mu_i} \Gamma(\mu_i)}$,  $B_i = \frac{\mu_i}{\Omega_i^{\alpha_i\mu_i}}$, $i=1,2,\cdots,N$, $\{\alpha_i, \mu_i, \Omega_i\}$ are the fading parameters for the $i$-th antenna of the RF  link.

We denote by $H_t$ the  path gain, by $h_t=h_ph_{ft}$ the fading coefficient (where $h_p$ models the pointing error and $h_{ft}$ models the short term fading), and $\sigma_{w_t}^2$ by noise variance of the THz link, respectively.   In the second hop, we employ the fixed-gain AF relaying to the combiner output  with  a gain $G$  resulting the end-to-end SNR of the RF-THz link as \cite{Hasna_2004_AF}:
\begin{eqnarray}
	\label{eq:af}
	\gamma = \frac{\gamma_{r}\gamma_{t}}{\gamma_{t}+C}
\end{eqnarray}
where $C={P}/G^2\sigma_{w_t}^2$ is the gain dependent on the first link, and  $\gamma_t=\bar{\gamma}_{\rm thz}|h_{ft}h_p|^2$, where   $\bar{\gamma}_{\rm thz}= \frac{P |H_{t}|^2}{\sigma_{w_t}^2}$ denotes the average SNR of the THz link.  The PDF of SNR for the THz link  can be represented in terms of Meijer's G-function \cite{Boulogeorgos_Analytical}:
\begin{eqnarray}\label{eq:pdf_hfp}
	f_{\gamma_{t}}(x) = \frac{A_t \phi x^{\alpha_t\mu_t-1}S_0^{-\alpha_t\mu_t} }{ \alpha_t }  G_{1,2}^{2,0} \Bigg( \frac{B_t x^{\alpha_t}}{S_o^{\alpha_t}} \Bigg| \begin{matrix} 1-\mu_t+\frac{\phi}{\alpha_t} \\ -\mu_t+\frac{\phi}{\alpha_t}, 0 \end{matrix} \Bigg)
\end{eqnarray}
 where $A_t = \frac{\alpha_t\mu_t^{\mu_t}}{\Omega^{\alpha_t\mu_t} \Gamma(\mu_t)}$ and $B_t = \frac{\mu_t}{\Omega^{\alpha_t\mu_t}}$, and $S_0$ and $\phi$ are parameters that determine the severity of misalignment between the transmitter and receiver. 

\section{Performance Analysis}
In this section, we analyze the statistical performance of the 	considered RF-THz transmission by deriving analytical expressions of PDF and CDF of the end-to-end SNR, outage probability, average BER, and ergodic capacity.

Using \eqref{eq:af}, the well-known PDF of end-to-end SNR of the fixed-gain  AF relayed system is given by
\begin{eqnarray} \label{eq:pdf_eqn_af}
	&f_\gamma(z) = \int_{0}^{\infty}  {f_{\gamma_r}\left(\frac{z(x + C)}{x}\right)} {f_{\gamma_t}(x)} \frac{x + C}{{x}} {dx}
\end{eqnarray}
where $f_{\gamma_t}(x)$ is given in \eqref{eq:pdf_hfp}  and $f_{\gamma_r}$ denotes the PDF of the EGC output SNR $\gamma_r$. We can use \eqref{eq:pdf_hf_rf} and  \eqref{eq:egc_snr_eqn},  and apply the theory of random variables to derive the exact PDF of SNR $\gamma_r$ in terms of  multivariate Fox's H-function \cite{Mathai_2010}. However, the authors in  \cite{Costa_2008_alpha_mu_sum} demonstrated that a single $\alpha$-$\mu$ distribution can accurately approximate the sum of $N$ $\alpha$-$\mu$ variates. Thus, the PDF of \eqref{eq:egc_snr_eqn} can be approximated as
\begin{equation} \label{eq:pdf_hf_rf_egc}
	f_{\gamma_{r}}(\gamma) =  \frac{A_r \gamma ^{\frac{\alpha_r\mu_r}{2}-1}} {2 {\bar{\gamma}_{\rm r}}^{\frac{\alpha_r\mu_r}{2}}} \exp \bigg(- B_r {\Big(\sqrt{\frac{\gamma}{\bar{\gamma}}_{r}}\Big)^{\alpha_r}}\bigg)
\end{equation}
where $A_r = \frac{\alpha_{r}\mu_r^{\mu_{r}}}{\Omega^{\alpha_{r}\mu_{r}} \Gamma(\mu_{r})}$,  $B_r = \frac{\mu_{r}}{\Omega^{\alpha_{r}}}$. Note that the method of moment matching is used to determine the $\alpha$-$\mu$ parameters  $\{\alpha_r, \mu_r, \Omega_r\}$ \cite{Costa_2008_alpha_mu_sum}. 

We present the PDF of SNR for the AF relaying in terms of bi-variate Fox's H-function in the following Theorem:
\begin{my_theorem}
Let $\phi$ and $S_0$ be the parameters of pointing errors of the THz link. If  $\alpha_t$,  $\mu_t$, and  $\alpha_r$,  $\mu_r$ are  the fading parameters of THz and RF links, respectively,  then the PDF of end-to-end SNR for the fixed-gain AF relay is approximated by 
\begin{eqnarray}\label{pdf_relay_fox}
	&f_{\gamma}(z) =  \frac{A_t A_r \phi S_0^{-\alpha_t\mu_t}  {\bar{\gamma}_{\rm thz}}^{-{\frac{\alpha_t\mu_t}{2}}} C^{\frac{\alpha_t\mu_t}{2}} {\bar{\gamma}_{\rm rf}}^{-\frac{\alpha_r\mu_r}{2}} {z}^{\frac{\alpha_r\mu_r-2}{2}}} {4\alpha_t} \nonumber \\ & H_{1,0:1,3:1,1}^{0,1:3,0:0,1} \left[ \frac{B_t {C_1}^{\frac{\alpha_t}{2}}}{S_0^{\alpha_t} {\bar{\gamma}_{\rm thz}}^{\frac{\alpha_t}{2}}}, \frac{ {\bar{\gamma}_{\rm rf}}^{\frac{\alpha_r}{2}}} {B_r z^{\frac{\alpha_r}{2}}}  \Bigg| \begin{matrix}~~~V_1~~~	 \\ ~~~V_2~~~   \end{matrix}    \right]
\end{eqnarray}
where $V_1 = \big\{\big(1-\big(\frac{\alpha_t\mu_t-\alpha_r\mu_r}{2}\big);\frac{\alpha_t}{2},\frac{\alpha_r}{2} \big)\big\} : \big\{ \big(1+\frac{\phi}{\alpha_t}-\mu_t,1 \big)\big\} : \big\{\big(1,1\big) \big\}$, and $ V_2 = \big\{ - \big\} : \big\{ \big(\frac{\phi}{\alpha_t}-\mu_t,1 \big),  \big(0,1\big), \big(\frac{-\alpha_t\mu_t}{2},\frac{\alpha_t}{2}\big) \big\}:\big\{\big(1+\frac{\alpha_r\mu_r}{2}, \frac{\alpha_r}{2}\big)\big\}$.
\end{my_theorem}

\begin{IEEEproof}
See Appendix A.
\end{IEEEproof}

\subsection{Outage Probability}
Outage probability of a system is the probability that the instantaneous SNR is less  than some threshold value $\gamma_{\rm th}$, i.e.  $ P(\gamma <\gamma_{th})$.  Thus, substituting \eqref{pdf_relay_fox} in $F_{\gamma}(\gamma_{}) = \int_{0}^{\gamma_{\rm th}} f_{\gamma}(z) {dz}$, and applying the definition of Fox's H function with simplification of the inner integral 
\begin{eqnarray}
 	\int_{0}^{\gamma_{\rm th}} z^{\frac{\alpha_r\mu_r-\alpha_rs_2-2}{2}} dz=\frac {\gamma_{\rm th}^{{\frac{\alpha_r\mu_r-\alpha_rs_2}{2}}} \Gamma\big(\frac{\alpha_r\mu_r+\alpha_rs_2}{2}\big)}{\Gamma\big(1+\frac{\alpha_r\mu_r+\alpha_rs_2}{2}\big)}
\end{eqnarray}
we get an approximation on the outage probability as 
\begin{eqnarray}\label{cdf_relay_fox}
	&F_{\gamma}(\gamma_{}) = \frac{A_t A_r \phi S_0^{-\alpha_t\mu_t}  {\bar{\gamma}_{\rm thz}}^{-{\frac{\alpha_t\mu_t}{2}}} C^{\frac{\alpha_t\mu_t}{2}} {\bar{\gamma}_{\rm rf}}^{-\frac{\alpha_r\mu_r}{2}} {\gamma_{\rm th}}^{\frac{\alpha_r\mu_r}{2}}} {4\alpha_t} \nonumber \\ & H_{1,0:1,3:2,2}^{0,1:3,0:1,1} \bigg[ \frac{B_t {C_1}^{\frac{\alpha_t}{2}}}{S_0^{\alpha_t} {\bar{\gamma}_{\rm thz}}^{\frac{\alpha_t}{2}}}, \frac{ {\bar{\gamma}_{\rm rf}}^{\frac{\alpha_r}{2}}} {B_r {\gamma_{\rm th}}^{\frac{\alpha_r}{2}}}  \bigg| \begin{matrix} ~~Q_1~~ \\  ~~Q_2~~ \end{matrix} \bigg]
\end{eqnarray}
where $Q_1 = \big\{\big(1-\frac{\alpha_t\mu_t-\alpha_r\mu_r}{2};\frac{\alpha_t}{2},\frac{\alpha_r}{2} \big)\big\} : \big\{ \big(1+\frac{\phi}{\alpha_t}-\mu_t,1 \big)\big\} : \big\{\big(1,1\big), \big(1+\frac{\alpha_r\mu_r}{2}, \frac{\alpha_r}{2}\big) \big\}$, and $ Q_2 = \big\{ - \big\} : \big\{ \big(\frac{\phi}{\alpha_t}-\mu_t,1 \big),  \big(0,1\big), \big(\frac{-\alpha_t\mu_t}{2},\frac{\alpha_t}{2}\big) \big\}:\big\{\big(\frac{\alpha_r\mu_r}{2}, \frac{\alpha_r}{2}\big), \big(1+\frac{\alpha_r\mu_r}{2}, \frac{\alpha_r}{2}\big)\big\}$.
 
We use  \cite[Th. 1.7, 1.11]{Kilbas_2004} and compute  residues of \eqref{cdf_relay_fox} for both contours $L_1$ and $L_2$ at poles $s_1=0$, $\frac{\phi}{\alpha_t}-\mu_t$, $-\mu_t$ and $s_2=0$, $\frac{\alpha_ts_1+\alpha_t\mu_t-\alpha_r\mu_r}{\alpha_r}$ to represent the asymptotic  outage probability in the high SNR regime as given in \eqref{eq:outage_asymptotic} (see top of the next page).
\begin{figure*}[h]
\begin{eqnarray}\label{eq:outage_asymptotic}
	&P_{\rm out}^{\infty} = \frac{A_tA_r \phi S_0^{-\alpha_t\mu_t}{\rm\gamma_{\rm th}}^{\frac{\alpha_t\mu_t}{2}}}{4\alpha_t} \Bigg[\Bigg(\frac{2\Gamma\big(\frac{\phi-\alpha_r\mu_r}{2}\big){\Gamma(\mu_t-\frac{\phi}{\alpha_t})\Gamma(-\frac{\phi}{2})} C^{\frac{\phi}{2}}} {{\alpha_r\mu_r}\Gamma(\frac{-\alpha_r\mu_r}{2}){\bar{\gamma}_t}^{\frac{\phi}{2}}}\Big(\frac{B_t }{{S_0}^{\alpha_t}}\Big)^{\frac{\phi}{\alpha_t}-\mu_t}+\frac{2\Gamma\big(\frac{\alpha_t\mu_t-\alpha_r\mu_r}{2}\big)\Gamma(\frac{\phi}{\alpha_t}-\mu_t)\Gamma(\frac{-\alpha_t\mu_t}{2})C^{\frac{\alpha_t\mu_t}{2}}{\bar{\gamma}_t}^{-\frac{\alpha_t\mu_t}{2}}}{\Gamma(1+\frac{\phi}{\alpha_t}){\alpha_r\mu_r}\Gamma(\frac{-\alpha_r\mu_r}{2})}\nonumber\\&+\frac{4\Gamma(\mu_t)}{\phi\alpha_r\mu_r}\Big(\frac{B_t}{{S_0}^{\alpha_t}}\Big)^{-\mu_t}+\frac{8\Gamma\big(\frac{\alpha_t\mu_t-\alpha_r\mu_r}{\alpha_t}\big)C^{\frac{\alpha_r\mu_r}{2}}}{\alpha_r(\phi-\alpha_r\mu_r)\alpha_r\mu_r{\bar{\gamma}_t}^{\frac{\alpha_r\mu_r}{2}}}\Big(\frac{B_t}{{S_0}^{\alpha_t}}\Big)^{\frac{\alpha_r\mu_r-\alpha_t\mu_t}{\alpha_t}}\Bigg){\bar{\gamma}_r}^{-\frac{\alpha_r\mu_r}{2}}\nonumber\\&+\Bigg(\frac{4\Gamma\big(\mu_t-\frac{\phi}{\alpha_t}\big)\Gamma\big(\frac{-\phi+\alpha_r\mu_r}{\alpha_r}\big){B_r}^{\frac{\phi}{\alpha_r}-\mu_r}C^{\frac{\phi}{2}}}{\alpha_r\phi{\bar{\gamma}_t}^\frac{\phi}{2}}\Big(\frac{B_t}{{S_0}^{\alpha_t}}\Big)^{\frac{\phi}{\alpha_t}-\mu_t}\Bigg)\big({\bar{\gamma}_r}\big)^{-\frac{\phi}{2}}+\Bigg(\frac{4\Gamma\big(\frac{-\alpha_t\mu_t+\alpha_r\mu_r}{\alpha_r}\big){B_r}^{\frac{\alpha_t\mu_t-\alpha_r\mu_r}{\alpha_r}}C^{\frac{\alpha_t\mu_t}{2}}}{\alpha_r(\frac{\phi}{\alpha_t}-\mu_t)\alpha_t\mu_t{\bar{\gamma}_r}^{\frac{\alpha_t\mu_t}{2}}}\Bigg){\bar{\gamma}_t}^{-\frac{\alpha_t\mu_t}{2}}\Bigg] 
\end{eqnarray}
	\hrule
\end{figure*}

Further, it is evident from \eqref{eq:outage_asymptotic} that diversity order for the outage probability of the system is $\min \big\{ \frac{\alpha_r \mu_r}{2},\frac{\alpha_t \mu_t}{2}, \frac{\phi}{2}\big\}$. It is worth mentioning that the diversity order of the fixed-gain AF is the same as that of the DF relaying \cite{Pranay_2021_TVT}. 	Using the property I from \cite{Costa_2008_alpha_mu_sum}, it can be seen that the resultant $\mu_r$ parameter increases with an increase in the number of antennas, $N$.  Thus, if the pointing error is sufficiently reduced by adjusting the signal beam-width, the overall system performance will be determined by the fading parameters of the THz channel.

\subsection{Average BER}\label{sec:av_ber}
The average BER of a communication system with binary modulation is given as:
\begin{eqnarray} \label{eq:ber}
	\bar{P}_e = \frac{q^p}{2\Gamma(p)}\int_{0}^{\infty} \gamma^{p-1} {e^{{-q \gamma}}} F_{\gamma} (\gamma)   d\gamma
\end{eqnarray}
where $p$ and $q$ are the constants that determine the type of modulation used. Using CDF of \eqref{cdf_relay_fox} in \eqref{eq:ber}, the average BER can be approximated as
\begin{eqnarray} \label{eq:ber_proof_int}
	&\bar{P}_e = \frac{A_tA_r \phi S_0^{-\alpha_t\mu_t} {\bar{\gamma}_{\rm thz}}^{-\frac{\alpha_t\mu_t}{2}} {\bar{\gamma}_{\rm rf}}^{-\frac{\alpha_r\mu_r}{2}}  C^{\frac{\alpha_t\mu_t}{2}} }{4\alpha_t} \Big(\frac{1}{2\pi \J}\Big)^2 \nonumber \\ &    \int_{L_1}^{} \int_{L_2}^{}      \frac{ \Gamma(\frac{\phi}{\alpha_t}-\mu_t-s_1) \Gamma(0-s_1)} {\Gamma(1+\frac{\phi}{\alpha_t}-\mu_t-s_1)} \Big(\frac{B_t C^{\frac{\alpha_t}{2}}}{{S_0}^{\alpha_t} {{\bar{\gamma}_{\rm thz}}^{\frac{\alpha_t}{2}}}} \Big)^{s_1} \Gamma(s_2) \big( \frac{ {\bar{\gamma}_{\rm rf}}^{\frac{\alpha_r}{2}}} {B_r}\big)^{s_2}   \nonumber \\ &   \frac{ \Gamma(\frac{-\alpha_t\mu_t -\alpha_ts_1}{2}) \Gamma(\frac{\alpha_t\mu_t-\alpha_r\mu_r +\alpha_ts_1+\alpha_rs_2}{2})}{\Gamma(\frac{-\alpha_r\mu_r+\alpha_rs_2}{2})}  ds_1 ds_2 \nonumber \\ &   \int_{0}^{\infty} {\gamma}^{\frac{\alpha_r\mu_r-\alpha_rs_2}{2}} \gamma^{p-1} e^{-q\gamma}  d{\gamma}
\end{eqnarray}
Using the solution of inner integral $\int_{0}^{\infty} {\gamma}^{\frac{\alpha_r\mu_r--\alpha_rs_2}{2}}  \gamma^{p-1} e^{-q\gamma}d{\gamma}$ [\cite{Gradshteyn}, eq.(3.381/4)] in terms of Gamma function, and applying the definition of Fox's H-function \cite[(1.1)]{Mittal_1972}, we get 
\begin{eqnarray}\label{ber_relay_fox}
	&\bar{P}_e = \frac{A_t A_r \phi S_0^{-\alpha_t\mu_t}  {\bar{\gamma}_{\rm thz}}^{-{\frac{\alpha_t\mu_t}{2}}} C^{\frac{\alpha_t\mu_t}{2}} {\bar{\gamma}_{\rm rf}}^{-\frac{\alpha_r\mu_r}{2}} q^{-\big(\frac{\alpha_r\mu_r}{2}+p\big)}} {4\alpha_t} \nonumber \\ & H_{1,0:1,3:2,3}^{0,1:3,0:2,1} \left[ \frac{B_t {C_1}^{\frac{\alpha_t}{2}}}{S_0^{\alpha_t} {\bar{\gamma}_{\rm thz}}^{\frac{\alpha_t}{2}}}, \frac{ {\bar{\gamma}_{\rm rf}}^{\frac{\alpha_r}{2}} q^{\frac{\alpha_r}{2}}} {B_r} \Bigg| \begin{matrix} ~~U_1~~ \\ ~~U_2~~  \end{matrix} \right]
\end{eqnarray}
where $U_1 = \big\{\big(1-\frac{\alpha_t\mu_t-\alpha_r\mu_r}{2};\frac{\alpha_t}{2},\frac{\alpha_r}{2} \big)\big\} : \big\{ \big(1+\frac{\phi}{\alpha_t}-\mu_t,1 \big)\big\} : \big\{\big(1,1\big), \big(1+\frac{\alpha_r\mu_r}{2}, \frac{\alpha_r}{2}\big) \big\}$, and $ U_2 = \big\{ - \big\} : \big\{ \big(\frac{\phi}{\alpha_t}-\mu_t,1 \big),  \big(0,1\big), \big(\frac{-\alpha_t\mu_t}{2},\frac{\alpha_t}{2}\big) \big\}:\big\{\big(\frac{\alpha_r\mu_r}{2}, \frac{\alpha_r}{2}\big), \big(p+\frac{\alpha_r\mu_r}{2}, \frac{\alpha_r}{2}\big) \big(1+\frac{\alpha_r\mu_r}{2}, \frac{\alpha_r}{2}\big)\big\} $. 
		
Upon following a similar analysis as the asymptotic expression of the outage probability, we can derive the average BER at the high SNR in terms of simpler Gamma function and the corresponding diversity order will be the same, i.e. $\min \big\{ \frac{\alpha_r \mu_r}{2},\frac{\alpha_t \mu_t}{2}, \frac{\phi}{2}\big\}$.

\subsection{Ergodic Capacity}\label{sec:capacity}
The ergodic capacity is defined as
\begin{eqnarray} \label{eq:capacity_eqn}
	\overline{C}=  \int_{0}^{\infty} {\rm log_2}(1+\gamma) f_\gamma(\gamma) d\gamma 
\end{eqnarray}
Thus, substituting the PDF \eqref{pdf_relay_fox} in \eqref{eq:capacity_eqn} and using the integral representation of Fox's H-function gives us the inner integral \cite[4.293.3]{Gradshteyn}
\begin{eqnarray}
\int_{0}^{\infty} {\rm log}(1+\gamma) {\gamma}^{\frac{\alpha_r\mu_r-\alpha_rs_2-2}{2}}  d{\gamma}= \frac{\pi Csc(\frac{\pi (\alpha_2\mu_2+\alpha_2S_2)}{2})}{{\rm log(2)} (\frac{\alpha_2\mu_2+\alpha_2S_2}{2})}
\end{eqnarray}
Finally, we simplify the inner integral using the identity $\pi Csc(\pi a) = \Gamma(a)\Gamma(1-a)$ and apply the definition of Fox's H-function \cite[1.1]{Mittal_1972}, to get an approximate expression of the ergodic capacity as 

\begin{eqnarray}\label{capacity_relay_fox}
	&\overline{C} =  \frac{ A_t A_r \phi S_0^{-\alpha_t\mu_t}  {\bar{\gamma}_{\rm thz}}^{-{\frac{\alpha_t\mu_t}{2}}} C^{\frac{\alpha_t\mu_t}{2}} {\bar{\gamma}_{\rm rf}}^{-\frac{\alpha_r\mu_r}{2}} } {4\alpha_t {\rm log}(2)} \nonumber \\ & H_{1,0:1,3:3,3}^{0,1:3,0:2,2} \bigg[ \frac{B_t {C_1}^{\frac{\alpha_t}{2}}}{S_0^{\alpha_t} {\bar{\gamma}_{\rm thz}}^{\frac{\alpha_t}{2}}}, \frac{ {\bar{\gamma}_{\rm rf}}^{\frac{\alpha_r}{2}}} {B_r}  \bigg| \begin{matrix}
	~~D_1~~	  \\ ~~D_2~~   \end{matrix}    \bigg]
\end{eqnarray}
where  $D_1 = \big\{\big(1-\frac{\alpha_t\mu_t-\alpha_r\mu_r}{2};\frac{\alpha_t}{2},\frac{\alpha_r}{2} \big)\big\} : \big\{ \big(1+\frac{\phi}{\alpha_t}-\mu_t,1 \big)\big\} : \big\{\big(1,1\big), \big(\frac{\alpha_r\mu_r}{2}, \frac{\alpha_r}{2}\big), \big(1+\frac{\alpha_r\mu_r}{2}, \frac{\alpha_r}{2}\big) \big\}$, and $ \big\{ - \big\} : \big\{ \big(\frac{\phi}{\alpha_t}-\mu_t,1 \big),  \big(0,1\big), \big(\frac{-\alpha_t\mu_t}{2},\frac{\alpha_t}{2}\big) \big\}:\big\{\big(\frac{\alpha_r\mu_r}{2}, \frac{\alpha_r}{2}\big), \big(\frac{\alpha_r\mu_r}{2}, \frac{\alpha_r}{2}\big), \big(1+\frac{\alpha_r\mu_r}{2}, \frac{\alpha_r}{2}\big)\big\} $.

\begin{figure*}[t]
	\subfigure[Outage probability.]{\includegraphics[scale=0.33]{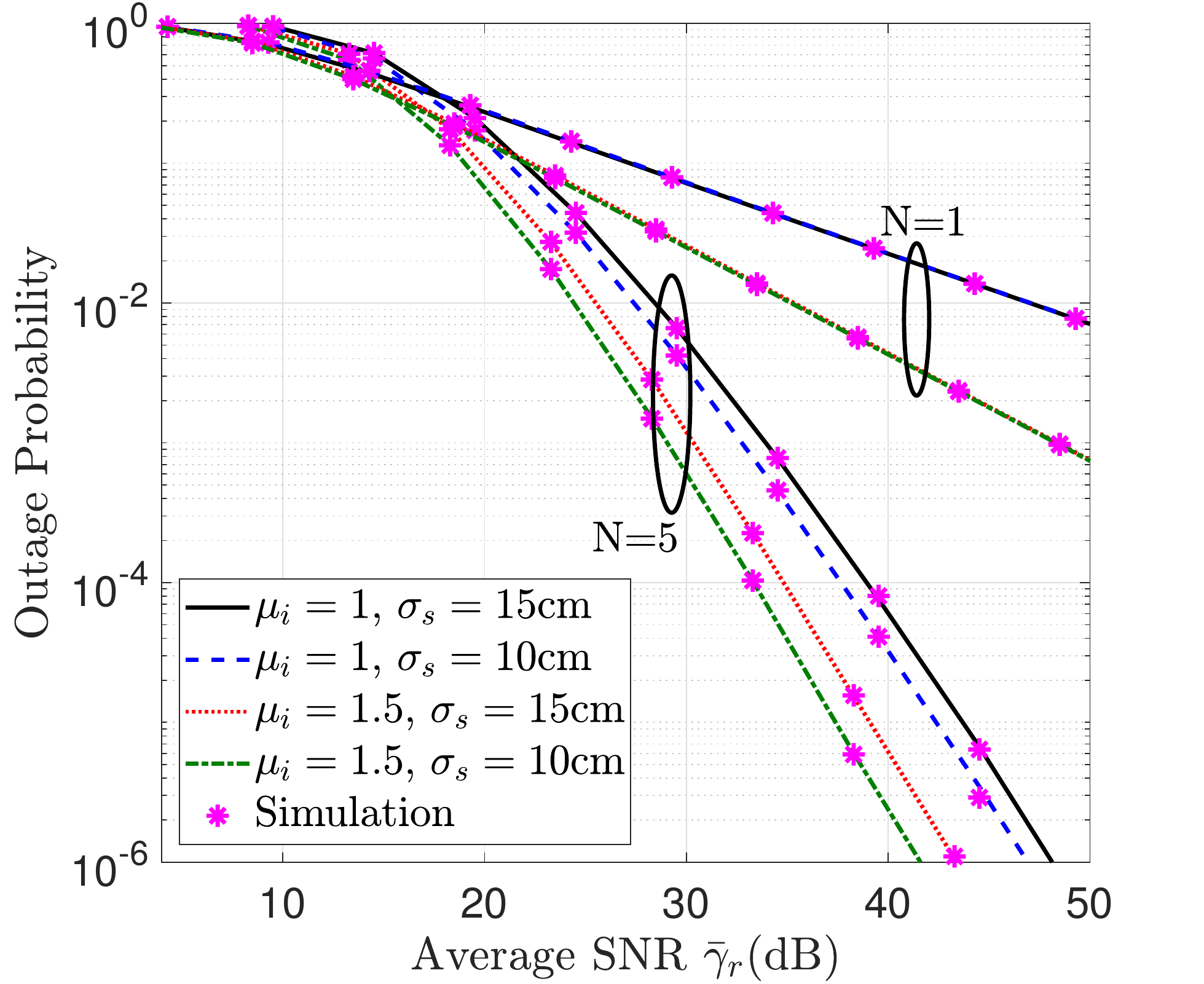}} \hspace{-5mm}
	\subfigure[{Average BER with $\mu_i=1.2$  and $\sigma_s=8 \mbox{cm}$.} ]{\includegraphics[scale=0.33]{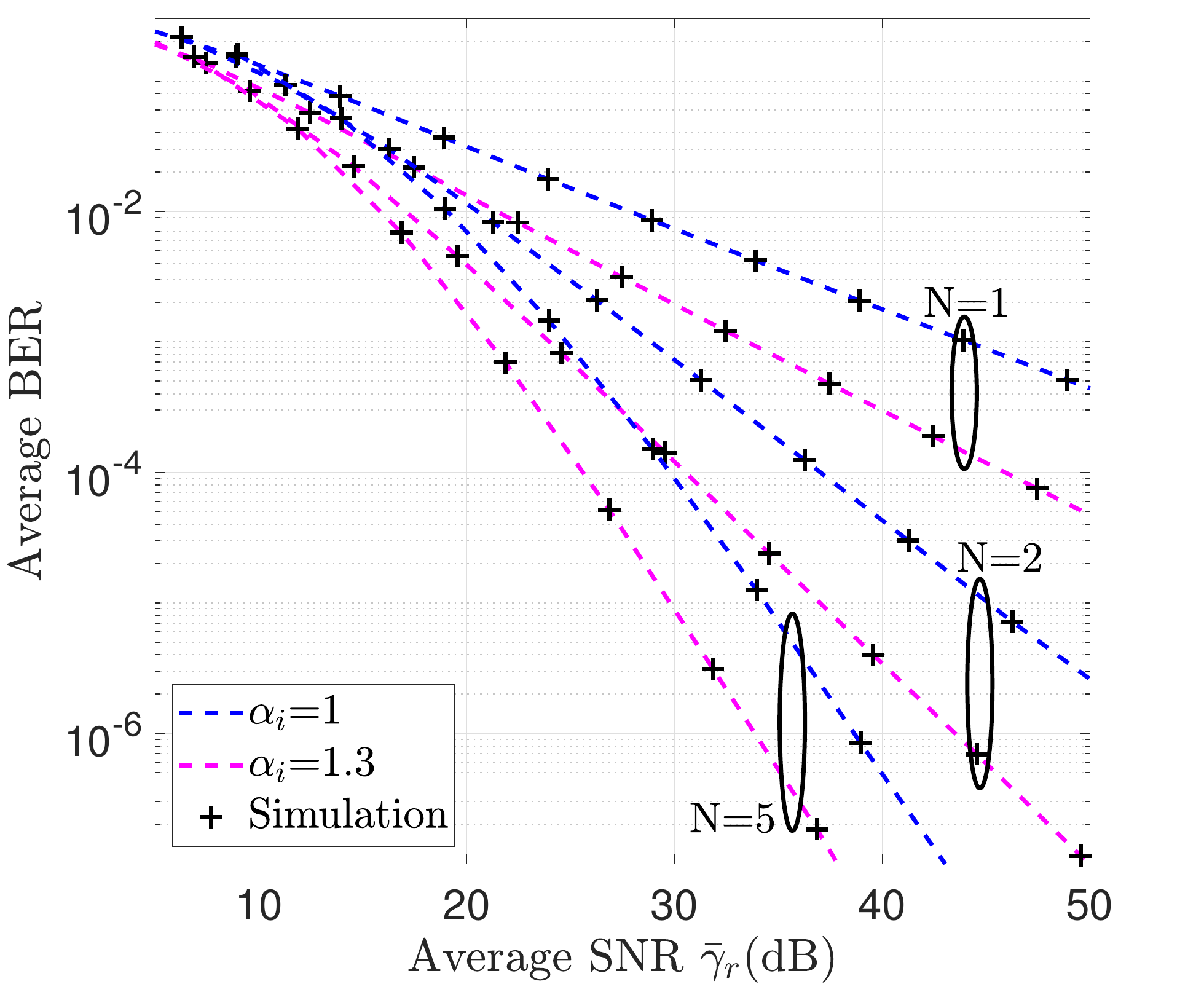}} \hspace{-5mm}
	\subfigure[Ergodic capacity for THz link distance $d_t=50$\mbox{m}. ]{\includegraphics[scale=0.33]{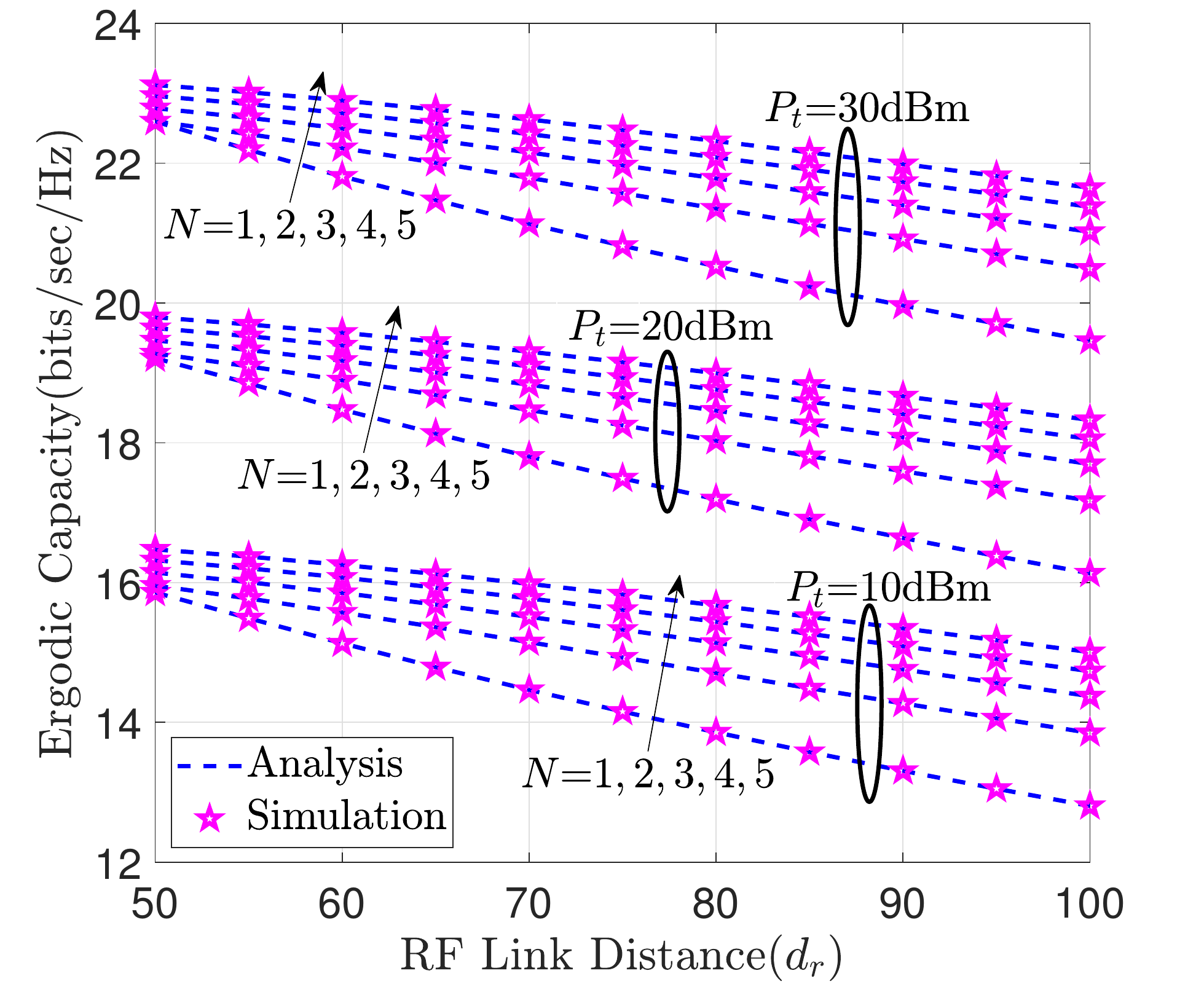}}
	\caption{Performance of the EGC-enabled RF mixed with THz wireless system with $\alpha_t=2$  and $\mu_t=2.6$.}
	\label{fig:outage_egc}
	\label{fig:ber_egc}
	\label{fig:capacity_egc}
\end{figure*}

\section{Simulation and Numerical Results}\label{sec:sim_results}
In this section, we validate the derived analytical results with the help of Monte-Carlo simulations averaged over $10^8$ iterations and evaluate the performance of the hybrid RF-THz system. We compute the path gain for RF link using 3GPP standard path loss model $L_r({\rm dB}) = 32.4+17.3\log_{10}(d_2)+20\log_{10} (10^{-9}f_r)$ for the RF link distance ($d_r$) varying from $50$\mbox{m}-$100$\mbox{m}. We use  $26$\mbox{dBi}  gain for each antenna for the RF link, carrier frequency $f_r$= $6$\mbox{GHz}, and RF signal bandwidth $20$\mbox{MHz}. The higher RF antenna gain is considered for illustration such that the performance of the RF-THz system depends on either  RF or THz links.  To compute the path gain of the THz link, we use $H_{t} = \frac{cG_t}{4\pi f_t d_t} \exp(-\frac{1}{2}kd_t)$, where $G_t=55$\mbox{dBi}, $k=2.8\times10^{-4}$ is the absorption coefficient \cite{Boulogeorgos_Error}, $c$ is the speed of light, $f_t= 0.275$ \mbox{THz}, THz bandwidth of $10$\mbox{GHz}, and THz link distance $d_t=50$\mbox{m}. We use MATLAB implementation to evaluate the bivariate Fox’s H-function, as given in \cite{Illi_2017}.

First, we demonstrate the impact of  fading parameters and pointing errors on the system's outage probability with an increase in the number of antennas for the RF link, as depicted in Fig. \ref{fig:outage_egc}(a). The outage performance of the system improves significantly (more than $100$ times) as the number of antennas is increased from $1$ to $5$ for $\mu_i=1$ at an average SNR of $40$ \mbox{dB}. Increasing the multi-path clustering parameter $\mu_i$ also improves the outage performance of the system as it improves the channel conditions.  The figure shows that the effect of pointing error is negligible for a single antenna case ($N=1$) since the outage probability is determined by the weaker of two links (here, the RF link), and thus plots corresponding to different $\sigma_s$ (pointing error parameter of the THz link) become indistinguishable. The diversity order $\min\{\frac{\alpha_1\mu_1}{2}, \frac{\alpha_t\mu_t}{2}, \frac{\phi}{2}\}$ i.e., $\min\{0.5,2.6,2\}=0.5$ also support this claim.   However, as the number of antennas increases from $N=1$ to $N=5$, the term $\frac{\alpha_{r}\mu_{r}}{2}$ increases due to the multi-path EGC diversity (see Property I of \cite{Costa_2008_alpha_mu_sum}).  Thus, the outage performance of the system  depends on the pointing error parameter $\phi$ of the THz link. Here, the diversity order becomes $\min\{2.5, 2.6, 2\}=2$ since $\phi$ is increased from $4$ to $9$ (corresponding to change in $\sigma_s$ from $15$ \mbox{cm} to $10$ \mbox{cm}). The change of slopes in the plots of Fig. \ref{fig:outage_egc}(a) confirms our diversity order analysis.

Next, we show the impact of the number of antennas and the non-linearity parameter $\alpha_i$ on the average BER of the mixed multi-antenna RF-THz system, as shown in Fig. \ref{fig:outage_egc}(b). The figure shows that the average BER reduces with an increase in the number of received RF antennas. Moreover, increasing $\alpha_i$ results in better average BER performance since the channel becomes more linear. The diversity order follows the same analysis as that of the outage probability, which can be confirmed with the  slope change among various plots. 

Finally, we evaluate the ergodic capacity performance of the mixed multi-antenna system by considering a fixed distance for THz link $d_t=50 \mbox{m}$ and different lengths of the RF link $d_r$ for different input power with $\alpha_i=1.5$, $\mu_i=1.5$ and $\sigma_s=5 \mbox{cm}$. The ergodic capacity decreases as the link distance increases and improves with the number of antennas at the AP.  The figure depicts a significant increase of $4$ \mbox{bits/sec/Hz} when  the input power is increased  from $10$ \mbox{dBm} to $20$ \mbox{dBm}.

\section{Conclusion}\label{sec:conc}
In this paper, we analyzed the performance of a hybrid dual-hop RF-THz system using fixed-gain AF protocol with multi-antenna EGC receiver for the RF transmission by considering  $\alpha$-$\mu$  channel fading for both the links and zero boresight pointing errors in THz link. We derived the PDF and CDF of the considered system in terms of bivariate Fox's H-function by approximating the PDF for the sum of $\alpha$-$\mu$ variates. We derived the outage probability, average BER, and ergodic capacity of the considered system and presented an asymptotic analysis of the outage probability in a high SNR region. The diversity order depicted that the multi-antenna at the AP improves the performance and that performance can be made dependent on the fading characteristics of the backhaul THz. We used computer simulations to demonstrate the scaling of performance with an increase in antennas. It would be interesting to investigate the performance of selection and maximal ratio combining receivers as future scope of the work.

\section*{Appendix A: PDF of the End-to-End SNR}
Representing the exponential function in terms of Meijer-G function, we represent the PDF of SNR for the RF link $f_{\gamma_r}(\gamma)$ in \eqref{eq:pdf_hf_rf_egc}:
\begin{eqnarray} \label{eq:pdf_hf_rf_MG}
	f_{\gamma_r}(\gamma) = \frac{A_r \big(\gamma^{\frac{1}{2}} {\bar{\gamma}_{\rm rf}}^{-\frac{1}{2}}\big)^{\alpha_r\mu_r-1}} {2\sqrt{\gamma\bar{\gamma}_{\rm rf}}}  G_{0, 1}^{1,0}  \Bigg( {B_r{\big(\gamma^{\frac{1}{2}} {\bar{\gamma}_{\rm rf}}^{-\frac{1}{2}}\big)}^{\alpha_r}} \Bigg| \begin{matrix} -\\0\end{matrix}\Bigg)                          
\end{eqnarray}
Substituting  \eqref{eq:pdf_hfp} and \eqref{eq:pdf_hf_rf_MG} in \eqref{eq:pdf_eqn_af}, using the integral representation of Meijer's G-function and changing the integral's order:
\begin{eqnarray} \label{eq:pdf_proof_int}
	&f_{{\gamma}}(z) = \frac{A_tA_r \phi S_0^{-\alpha_t\mu_t} {\bar{\gamma}_{\rm thz}}^{-\frac{\alpha_t\mu_t}{2}} {\bar{\gamma}_{\rm rf}}^{-\frac{\alpha_r\mu_r}{2}} z^{\frac{\alpha_r\mu_r-2}{2}} }{4\alpha_t}  \nonumber \\ &   \Big(\frac{1}{2\pi \J}\Big)^2 \int_{L_1}^{} \int_{L_2}^{}  \frac{ \Gamma(\frac{\phi}{\alpha_t}-\mu_t-s_1) \Gamma(0-s_1)} {\Gamma(1+\frac{\phi}{\alpha_t}-\mu_t-s_1)} \Big(\frac{B_t}{{S_0}^{\alpha_t} {{\bar{\gamma}_{\rm thz}}^{\frac{\alpha_t}{2}}}} \Big)^{s_1} \nonumber \\ &    \Gamma(0-s_2) \big( \frac{B_r z^{\frac{\alpha_r}{2}}} {{\bar{\gamma}_{\rm rf}}^{\frac{\alpha_r}{2}}}\big)^{s_2}  {\rm d}s_1 {\rm d}s_2 I_{in}
\end{eqnarray}
where $L_1$ and $L_2$ denote the contours of line integrals. We use \cite[3.194/3]{Gradshteyn} and \cite[8.384/1]{Gradshteyn}  to solve the inner integral $I_{in}$ in terms of the compatible Gamma function:
\begin{eqnarray} \label{inner_int_pdf}
	&I_{in} = \int_{0}^{\infty}  {x^{(\frac{\alpha_t\mu_t+\alpha_ts_1-2}{2})}} \Big(\frac{x+c}{x}\Big)^{(\frac{\alpha_r\mu_r+\alpha_rs_2}{2})} dx \nonumber \\ &  = \frac{C^{(\frac{\alpha_t\mu_t+\alpha_ts_1}{2})}  \Gamma(\frac{-\alpha_t\mu_t -\alpha_ts_1}{2}) \Gamma(\frac{\alpha_t\mu_t-\alpha_r\mu_r +\alpha_ts_1-\alpha_rs_2}{2})}{\Gamma(\frac{-\alpha_r\mu_r-\alpha_rs_2}{2})}
\end{eqnarray}
Finally, we substitute \eqref{inner_int_pdf} in \eqref{eq:pdf_proof_int}, and  to apply the definition of Fox's H-function \cite[1.1]{Mittal_1972}, we use  $s_2 \to -s_2 $ in \eqref{eq:pdf_proof_int} to get \eqref{pdf_relay_fox} of  Theorem 1.
\bibliographystyle{IEEEtran}
\bibliography{Thz_references_others}

\begin{thebibliography}{10}
\providecommand{\url}[1]{#1}
\csname url@samestyle\endcsname
\providecommand{\newblock}{\relax}
\providecommand{\bibinfo}[2]{#2}
\providecommand{\BIBentrySTDinterwordspacing}{\spaceskip=0pt\relax}
\providecommand{\BIBentryALTinterwordstretchfactor}{4}
\providecommand{\BIBentryALTinterwordspacing}{\spaceskip=\fontdimen2\font plus
\BIBentryALTinterwordstretchfactor\fontdimen3\font minus
  \fontdimen4\font\relax}
\providecommand{\BIBforeignlanguage}[2]{{%
\expandafter\ifx\csname l@#1\endcsname\relax
\typeout{** WARNING: IEEEtran.bst: No hyphenation pattern has been}%
\typeout{** loaded for the language `#1'. Using the pattern for}%
\typeout{** the default language instead.}%
\else
\language=\csname l@#1\endcsname
\fi
#2}}
\providecommand{\BIBdecl}{\relax}
\BIBdecl

\bibitem{Elayan_2019}
H.~{Elayan} \emph{et~al.}, ``Terahertz band: The last piece of {RF} spectrum
  puzzle for communication systems,'' \emph{IEEE Open J. Commun. Soc.}, vol.~1,
  pp. 1--32, 2020.

\bibitem{Koenig_2013_nature}
S.~{Koenig} \emph{et~al.}, ``Wireless {sub-THz} communication system with high
  data rate,'' \emph{Nature Photon}, vol.~7, p. 977–981, 2013.

\bibitem{Kokkoniemi_2018}
J.~{Kokkoniemi} \emph{et~al.}, ``Simplified molecular absorption loss model for
  275–400 {Gigahertz} frequency band,'' in \emph{12th Eur. Conf. Antennas
  Propag. (EuCAP)}, Apr. 2018, pp. 1--5.

\bibitem{Boulogeorgos_Error}
A.~A. {Boulogeorgos} and A.~{Alexiou}, ``Error analysis of mixed {THz-RF}
  wireless systems,'' \emph{IEEE Commun. Lett.}, vol.~24, no.~2, pp. 277--281,
  2020.

\bibitem{Pranay_2021_TVT}
P.~Bhardwaj and S.~M. Zafaruddin, ``Performance of dual-hop relaying for
  {THz-RF} wireless link over asymmetrical $\alpha$-$\mu$ fading,'' \emph{IEEE
  Trans. Veh. Technol.}, vol.~70, no.~10, pp. 10\,031--10\,047, 2021.

\bibitem{Pranay_2021_Letters}
------, ``Fixed-gain af relaying for {RF-THz} wireless system over
  $\alpha$-$\kappa$-$\mu$ shadowed and $\alpha$-$\mu$ channels,'' \emph{IEEE
  Commun. Lett.}, pp. 1--1, 2021.

\bibitem{Lee_2011_FSO_RF}
E.~Lee \emph{et~al.}, ``Performance analysis of the asymmetric dual-hop relay
  transmission with mixed {RF/FSO} links,'' \emph{IEEE Photon. Technol. Lett.},
  vol.~23, no.~21, pp. 1642--1644, 2011.

\bibitem{Ashrafzadeh_2019_FSO_RF}
B.~Ashrafzadeh \emph{et~al.}, ``A framework on the performance analysis of
  dual-hop mixed {FSO-RF} cooperative systems,'' \emph{IEEE Trans. on Commun.},
  vol.~67, no.~7, pp. 4939--4954, 2019.

\bibitem{Papasotiriou2021}
E.~N. Papasotiriou \emph{et~al.}, ``A new look to {THz} wireless links: Fading
  modeling and capacity assessment,'' in \emph{2021 IEEE 32nd Annu. Int. Symp.
  on Personal, Indoor and Mobile Radio Commun.}, 2021, pp. 1--6.

\bibitem{Mathai_2010}
A.~M.{ Mathai} \emph{et~al.}, ``The {H}-function: Theory and applications.''
  vol. New York, NY, USA, Springer, 2010.

\bibitem{Annamalai_2000_EGC}
A.~Annamalai \emph{et~al.}, ``Equal-gain diversity receiver performance in
  wireless channels,'' \emph{IEEE Trans. on Commun.}, vol.~48, no.~10, pp.
  1732--1745, 2000.

\bibitem{Hasna_2004_AF}
M.~Hasna and M.-S. Alouini, ``A performance study of dual-hop transmissions
  with fixed gain relays,'' \emph{IEEE Trans. Wireless Commun.}, vol.~3, no.~6,
  pp. 1963--1968, 2004.

\bibitem{Boulogeorgos_Analytical}
A.~A. {Boulogeorgos} and A.~{Alexiou}, ``Analytical performance assessment of
  {THz} wireless systems,'' \emph{IEEE Access}, vol.~7, pp. 11\,436--11\,453,
  2019.

\bibitem{Costa_2008_alpha_mu_sum}
D.~Costa \emph{et~al.}, ``Highly accurate closed-form approximations to the sum
  of $\alpha$-$\mu$ variates and applications,'' \emph{IEEE Trans. Wireless
  Commun.}, vol.~7, no.~9, pp. 3301--3306, 2008.

\bibitem{Kilbas_2004}
A.~A. {Kilbas}, \emph{H-Transforms: Theory and Applications}.\hskip 1em plus
  0.5em minus 0.4em\relax CRC Press, 2004, vol. First edition.

\bibitem{Gradshteyn}
I.~S. {Gradshteyn} and I.~M. {Ryzhik }, \emph{Table of Integrals, Series, and
  Products}.\hskip 1em plus 0.5em minus 0.4em\relax Academic press, San Diego,
  CA, 6th edition, 2000.

\bibitem{Mittal_1972}
P.~{Mittal} and K.~{Gupta}, ``An integral involving generalized function of two
  variables,'' \emph{Proc. Indian Acad. Sci.}, vol.~75, no.~9, pp. 117--123,
  1972.

\bibitem{Illi_2017}
E.~Illi \emph{et~al.}, ``A performance study of a hybrid {5G RF/FSO}
  transmission system,'' in \emph{2017 Int. Conf. on Wireless Netw. and Mobile
  Commun. (WINCOM)}, 2017, pp. 1--7.

\end{thebibliography}

\end{document}